\documentclass[conference]{IEEEtran}

\usepackage{amsmath, amssymb, amsfonts}
\usepackage{graphicx}
\usepackage{hyperref}
\usepackage{booktabs}
\usepackage{algorithm}
\usepackage{caption}
\usepackage{geometry}
\usepackage{enumitem}
\usepackage{color}
\usepackage{dirtytalk}
\usepackage{algorithmic}
\usepackage{hyperref}
\usepackage{authblk}

\newcommand{\Pfunc}[3]{\ensuremath{P(#1, #2, #3)}}
\newcommand{\trajectory}{\ensuremath{\tau}}
\newcommand{\stateS}{\ensuremath{s}}

\newcommand{\operator}{\ensuremath{h}}

\newcommand{\DronesNum}{\ensuremath{n}}

\newcommand{\adviceNum}{\ensuremath{k}}

\newcommand{\action}{\ensuremath{a}}

\newcommand{\AG}{AG}
\newcommand{\RM}{RM}
\newcommand{\detectiothreshold}{\ensuremath{hc}}
\newcommand{\pausetreshold}{\ensuremath{lc}}

\newcommand{\wights}{\ensuremath{w}}
\newcommand{\velocity}{\ensuremath{vel}}
\newcommand{\altitude}{\ensuremath{alt}}
\newcommand{\UFunc}{\ensuremath{U}}
\newcommand{\Model}{\ensuremath{M}}
\newcommand{\parametersForU}{\Psi}
\newcommand{\paramForU}{\psi}

\title{Advising Agent for Supporting Human-Multi-Drone Team
Collaboration}

\begin{document}
  
\author[ ]{Hodaya Barr}
\author[ ]{Dror Levy}
\author[ ]{Ariel Rosenfeld}
\author[ ]{Oleg Maksimov}
\author[ ]{Sarit Kraus}
\affil[ ]{Department of Computer Science, Bar-Ilan University, Ramat Gan, Israel}
\affil[ ]{\texttt{odayaben@gmail.com, 
dror10392@gmail.com,
ariel.rosenfeld@biu.ac.il}}
\affil[ ]{\texttt{oleg@maksimov.co.il,
sarit@cs.biu.ac.il}}

\maketitle
\begin{abstract}
Multi-drone systems have become transformative technologies across various industries, offering innovative applications. However, despite significant advancements, their autonomous capabilities remain inherently limited. As a result, human operators are often essential for supervising and controlling these systems, creating what is referred to as a human-multi-drone team. In realistic settings, human operators must make real-time decisions while addressing a variety of signals, such as drone statuses and sensor readings, and adapting to dynamic conditions and uncertainty. This complexity may lead to suboptimal operations, potentially compromising the overall effectiveness of the team. In critical contexts like Search And Rescue (SAR) missions, such inefficiencies can have costly consequences.
This work introduces an advising agent designed to enhance collaboration in human-multi-drone teams, with a specific focus on SAR scenarios. The advising agent is designed to assist the human operator by suggesting contextual actions worth taking. To that end, the agent employs a novel computation technique that relies on a small set of human demonstrations to generate varying realistic human-like trajectories. These trajectories are then generalized using machine learning for fast and accurate predictions of the long-term effects of different advice.
Through human evaluations, we demonstrate that our approach delivers high-quality assistance, resulting in significantly improved performance compared to baseline conditions.

\end{abstract}

\section{Introduction}
\label{Introduction}

Multi-drone systems are revolutionizing various industries including agriculture \cite{tsouros2019review}, construction \cite{sai2022survey}, and search and rescue (SAR) \cite{lyu2023unmanned}, to name a few \cite{hassanalian2017classifications}. 
As drone technology advances, the potential for innovative applications and their societal impact is expected to continue to grow \cite{tezza2019state}. Nevertheless, there are some inherent limitations to what multi-drone systems can handle autonomously. As such, it is common for human operators to provide an extra layer of monitoring, decision-making, and situational awareness, thus forming a so-called \say{human-multi-drone team} \cite{rebensky2022teammates,francos2023role}.  

Supervising and controlling multiple drones in dynamic and uncertain environments is a highly challenging task~\cite{nischal2024challenges}. Model-driven approaches have been proposed to facilitate human-on-the-loop drone missions, where autonomous decisions are made by drones but guided by human oversight to improve mission success in time-critical operations \cite{agrawal2020modeldriven}. In particular, the human operator is typically expected to monitor and respond to the drones' locations, statuses, sensor readings, and various other signals and information in a timely fashion while managing unexpected events, emergencies, and changing circumstances. 
Research by \cite{porat2016supervising} demonstrated that when control tasks are extensive, human operators can effectively collaborate with two drones. However, even experienced operators encountered difficulties managing three drones (or more) in complex settings, leading to a reduction in the overall team performance.

The integration of intelligent agents into mixed teams is becoming increasingly prevalent, offering new opportunities and challenges in team dynamics \cite{HumanAgent_Team_Dynamics,natarajan2024mixedinitiative}. Aligned with this important line of work, in this paper, we develop and evaluate a novel advising agent for supporting human-multi-drone team collaboration for lifelike tasks. 
In our setting, advice refers to action recommendations aimed at improving the overall team performance, which the advising agent generates based on contextual information. The presented agent relies on machine learning estimations of the long-term effects of various human-like actions which, in turn, are translated into useful advice. 
Central to our approach is the generation of \textit{realistic} trajectories that mimic human operators, using very few human demonstrations. Specifically, in order to adequately utilize machine learning to estimate the long-term effects of different actions, we propose a method to simulate the trajectories resulting from (unseen) human-like actions from a given setting using a limited number of human-generated examples. 
We instantiate our approach to a realistic SAR simulation, and through an extensive human study, we show that our agent is capable of providing high-quality advice, which, in turn, brings about favorable team performance compared to baseline conditions.
\section{Related Work}

Various techniques have been proposed to assist human operators in managing multiple drones.  Roughly speaking, these can be divided into two groups: Intelligent User Interfaces (IUIs) and intelligent advising agents. The former group typically seeks to  assist the operator in gaining a better situational awareness and understanding and providing more efficient command and control capabilities. Several prior works explore natural user interfaces, such as speech, gestures, body position, and eye-tracking, to facilitate intuitive drone interaction (e.g., \cite{fernandez2016natural, implementation2020Brandon, natural2022marina}). \cite{agrawal2021explaining} examines how different IUI design choices affect human situational awareness, while \cite{chen2022multi} proposes a task-based graphical interface to enhance control and situational awareness. Similarly, \cite{kostenko2022supervised} explores classifying operator functional states using physiological data to dynamically improve human-machine interaction in drone swarm piloting.
Boggess et al. \cite{boggess2023explainable,boggess2022toward} propose XAI methods to explain the behavior and decision-making processes of a team of robots to a human collaborator.


Previous research has introduced methods to improve interaction between a single automated team member (e.g., a robot) and a human operator to enhance overall collaboration \cite{edgar2023humanrobot,schleibaum2024adesse,azaria2015,azaria2012}. Projects like HADRON have developed AI-driven control systems to reduce human workload by improving tasks such as target and defect detection \cite{casado2024hadron}. \cite{al2020generating} proposed a forward simulation-based alert system to notify supervisors of potential negative events. In \cite{wu2024hierarchical}, natural language processed by LLMs allows human operators to communicate with a drone team, though no human studies were conducted, and timely comment generation was not evaluated.

We focus on intelligent advising agents that seek to support the operator by advising the operator on which actions to take  \cite{rosenfeld2017intelligent}. This approach complements the previously discussed methods, as it is interface-agnostic (i.e., capable of being integrated with any user interface) and independent of the AI tools used by the drones.  It is shown to be highly effective (i.e., better team performance).
The intelligent advising agents on which actions to take approach was explored in several domains; for example, \cite{trabelsi2023advice} provides an intelligent agent for the teleoperation of autonomous vehicles, and \cite{vered2020demand} developed an intelligent agent for a multi-vehicle setting in order to study ways to increase user trust. 

Most closely related to our work is  \cite{rosenfeld2017intelligent}, who tackled a similar challenge in a non-aerial environment. The authors suggested the use of a so-called \say{utopic environment} where they simulate the effects of different actions by assuming their ground robots operate optimally, never malfunction, and require no supervision or control from the human operator. In other words, the authors assume that the utopically-simulated long-term effects of an action are a reasonable approximation of the actual long-term effects. The rationale for using such a utopic environment is clear -  it makes the collection and annotation of extensive human demonstrations unnecessary, as these simulations execute fully autonomously. Unfortunately, this assumption is highly unrealistic in complex real-world environments, as noted by the original authors themselves. In particular, when considering multiple drones deployed in unexpected dynamic conditions that require human intervention, this oversimplifying assumption may lead to poor estimations and sub-optimal advice provision. In this work, we propose a new method to overcome this limitation by relaying on a small set of human demonstrations.

\section{Advice Provision}
\label{sec:formal}

\subsection{Formalization}
Next, we formally introduce the advice provision problem as a Markov Decision Problem (MDP) $<S,A,P,R,\gamma>$ \cite{Puterman1994}.

Let us consider a set of $n$ semi-autonomous drones engaged in a cooperative task, supervised and controlled by a single human operator, $\operator$.
The state space $S$ consists of all contextual information regarding the drones (e.g., status, altitude) and the task's and operator's domain-specific characteristics (e.g., time elapsed, performance). The operator can perform an action $a\in A$ during the task, at any time $t\in [0,\ldots T]$. Since $\operator$ can choose to execute no action at any given $t$, $NULL\in A$. In addition, not every action is possible at each state, and therefore, we denote $A(s)$ as the set of applicable actions at state $s$.
Let $\Pfunc{\stateS}{\action}{\stateS'} \ $
be the transition function that denotes the probability of transitioning from state $\stateS$ to state $\stateS'$ following action $a$.
Clearly, for each $\stateS \in S$, 
$\sum_{s' \in S} \Pfunc{\stateS}{\action}{\stateS'}=1$.
Note that, when $\action = NULL$, it holds that $\stateS \not= \stateS'$ since the environment is dynamically changing.  
Finally, let $R(s)$ be the domain-specific reward function,
and $\gamma\in[0,1]$ be the discount factor.
Importantly, performing an action takes time, depending on the operator's ability. The transition function and the cost function are unknown to the advising agent, yet they can be estimated through observations. 

Advice is guidance provided by an agent to the human operator as an action that the operator should take ($a \in A$). 
Crucially, the operator maintains the autonomy to assess the advice, consider alternative actions, and make a final decision (i.e., the advice is non-binding). At any point in time $t$, the agent may provide advice according to its advising policy $\pi$, a mapping from states to actions.
In an ideal setting, we would like the operator to follow an optimal policy, $\pi^*_\operator: S \rightarrow A$, that maximizes the expected accumulative future reward. However, the dynamic and uncertain environment causes the underlying optimization problem to be intractable and thus, an optimal policy cannot be computed in a reasonable time. To overcome this limitation, we propose a methodology for computing high-quality advice without explicitly deriving an optimal policy. 

\subsection{Proposed Methodology}

We propose an advice provision methodology consisting of two phases: an offline phase which is targeted at estimating the expected long-term reward of performing an action (i.e., reward estimation) using limited human demonstrations and a machine learning model; and an online phase, primarily designed for providing a suitable advice in a given state.

\noindent{\bf  Reward Estimation Model}

\begin{algorithm}[hbpt!]
\caption{Reward Estimation}\label{alg:generation}
\begin{algorithmic}[1]
\REQUIRE $k,\theta,T,D$
\STATE $D_{syn} \gets \emptyset$
\FORALL{$\textit{d} \in D$}   
        \STATE Estimate $\Tilde{C}_h(a)$ based on $d$
	\FOR{$i \gets 1$ to $k$}
        \STATE $t\gets 0$ \ \  $\trajectory \gets \emptyset$
        \ \ \ $s_0 \gets d_0[0]$ \label{S0}
        \WHILE{$t < T$}
            \STATE $A_t \gets A(s_t)$ 
            \STATE $a_t \gets Pertubate(A_t,d,t)$ \label{Pertubate}
             \STATE $s_{t+\Tilde{C}_h(a)} \gets GetNewState(s_t,a_t)$ \label{GetNewState}
             \STATE $t\gets t+\Tilde{C}_h(a)$
             \STATE $\trajectory \gets \trajectory \cdot \{(s_t,a_t)\}$
        \ENDWHILE
        \STATE $D_{syn}\gets D_{syn}\cup\trajectory$
\ENDFOR
        \ENDFOR
        \IF {$QualityAssurance(D_{syn},D,\theta)$} \label{QualityAssurance}
            \STATE $M\gets Train(D_{syn},D)$
            \RETURN $M$
        \ELSE
            \RETURN $Unsuccessful$
        \ENDIF


\end{algorithmic}
\end{algorithm}

As noted before, a central component in solving the underlying optimization problem is the proper estimation of the expected benefit from performing an action. To that end, we assume that a set of demonstrations is available, $D$,  where $d\in D$ is a trajectory $\trajectory$ consisting of state-action pairs, $((s_0,a_0),...,(s_T,a_T))$, denoting the actions that the operator took at each time and state. If $D$ is \say{sufficiently large and comprehensive}, then a machine learning model could be readily trained based on these demonstrations to approximate the effects of performing an action at a given state and time. However, as noted before, collecting such a set of demonstrations is typically highly expensive and time-consuming. Therefore, we assume $D$ is small and needs to be extended before it can be effectively used for training a machine learning model. As detailed in Algorithm \ref{alg:generation}, we propose the generation of high-quality \textit{synthetic} trajectories that mimic the real demonstrations in an offline fashion (i.e., before actual deployment of the advising agent). 
In words, we start by estimating the time it takes for a given operator to perform different actions using the entire trajectory, $\Tilde{C}_h(a)$ (e.g., using an average). Then, we go through a real trajectory ($d\in D)$ and pertubate it to create slight changes (e.g., choosing another action at a given probability). The action is then applied in a simulated environment using the transition function (e.g., the transition can also consider the operator's performance), a new state is reached and the process is repeated. Overall, $k$ synthetic trajectories are generated based on each real one. 
The resulting set of generated trajectories is then checked to ensure that it, indeed, closely mimics the real ones (i.e., quality assurance using statistical measures). If so, a machine learning model is trained on the \textit{extended} set of demonstrations (otherwise, an \say{unsuccessful} message is raised). 
The trained model is utilized next in the online phase to rank the various actions.

\begin{figure}[t]
\centering
\includegraphics[width=0.9\columnwidth]{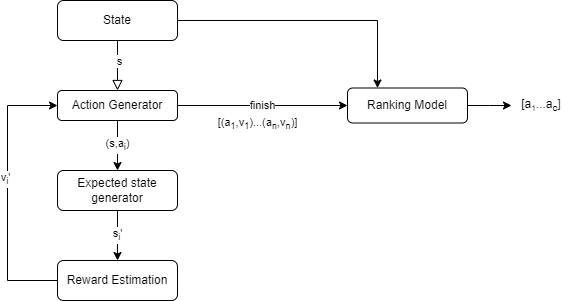}
\caption{The advising agent design. $s$ denotes the state, $a_i\in A$ denotes an action, $s'$ denotes the expected resulting state and $v_i$ denotes the predicted reward from the transition.}
\label{fig:agent_structure}
\end{figure}

\noindent{\bf Advice Provision}
During deployment, the agent employs an online advice provision policy as follows:  \\
{1. \bf Action generator}: Given state $s$, the agent generates a set of possible actions that can be performed at this state and time point. These actions may be the result of a drone-initiated event (e.g., a drone malfunction) and/or other actions aimed at improving performance (e.g., changing a drone's altitude for better coverage). It then evaluates them using the reward estimation model obtained in the offline phase (i.e., Algorithm \ref{alg:generation}). \\
{2. \bf Ranking}: All generated actions and their associated expected reward are provided to a ranking model that outputs the top $c$ actions as advice for the operator. See Figure \ref{fig:agent_structure} for an illustration. 
 
In other words, once a new state is encountered, the action generator creates possible actions to take which are then evaluated using the offline trained model and ranked accordingly.   



\section{Search and Rescue}
\label{sec:sar_mission}

We concentrate on the multidrone-based SAR task with a single human operator supervising and controlling a fleet of $\DronesNum$ drones in search of an unknown number of targets in a large area (e.g., searching for victims after an earthquake). The use of multiple drones working together as a team can reduce the time required to locate persons in distress, offering significant advantages over traditional methods \cite{hoang2023droneswarms}.  The typical overarching goal of such tasks, as in our setting, is to maximize the number of targets found during the mission's fixed time.
For a comprehensive review of the current state and challenges of drone applications in disaster management, including search and rescue, see~\cite{daud2022applications}.

Mission planning for multiple drones involves complex strategies to ensure efficient coverage and task completion \cite{song2023survey}.
We build on top of the drone-based SAR task modeling provided by \cite{du2019evolutionary}. Specifically, we assume that the entire search zone is known in advance and is partitioned into smaller sub-areas, denoted by $E$ the set of the sub-areas. Each sub-area is assigned a probability indicating the likelihood of finding at least one target within that specific sub-area. Assuming no prior knowledge, all sub-areas are given the same likelihood. Nonetheless, these probabilities may change dynamically based on the real-time information provided by the drones and/or manually by the operator (e.g.,  based on gathered intelligence that occasionally provided to the user through \say{intelligence messages}). In turn, these probabilities can influence the operator's decision to assign drones to specific sub-areas.

In our environment, drones scan their designated sub-areas following a lawn mower pattern. The scanning process is governed by four parameters - the drone's velocity, altitude, and two thresholds to determine the minimal confidence for a low-confidence suspected target and a high-confidence suspected target. Specifically, during the scanning phase, the drones' cameras capture images rapidly. These images are then processed through a neural network (NN) to produce bounding boxes and associated confidence indicating the presence of potential targets. To that end, a Retina net \cite{retina:20} NN model is trained using the approach provided in \cite{AIR:21} on the Heridal dataset \cite{heridal-lrbkc_dataset} and subsequently fine-tuned using manually tagged images from our simulated environment. It was shown that drones have difficulties identifying victims in SAR in the wild. Manzini and  Murphy \cite{manzini2023open} demonstrated that despite achieving performance that is statistically equivalent to the state-of-the-art on benchmark datasets, the models they tested fail to translate these achievements to the real world in terms of many false positives (e.g., identifying tree limbs and rocks as people), and false negatives (e.g., failing to identify victims). Similar problems were observed in our SAR environment in preliminary testing. 
Therefore, in our environment, the human operator needs to approve a suspected target. To balance between false alarms and missed detections, we define two thresholds: low-confidence (\pausetreshold) and high-confidence (\detectiothreshold). If the confidence of the NN exceeds \detectiothreshold, the drone stops, and an alert with the associated confidence is sent to the operator. If the confidence only surpasses the \pausetreshold, the drone further scans around the suspected object, and an alert with a timeout and the (relatively low) confidence of the detection is sent to the operator, allowing her to observe the object if she has the time. If no high confidence is achieved during further scanning, the drone dismisses the detection; otherwise, it elevates the alert to a high-confidence one and waits for the operator's handling of the event.
The drone's altitude, velocity, and the two confidence thresholds are determined by the operator based on the area's characteristics, existing intelligence, and the operator's capabilities, to name a few.
For example, highly dense sub-areas (e.g., forest) may be better assigned different parameters than sparse sub-areas (e.g., desert). The operator can determine the parameters of a given sub-area by designating its so-called \say{area type} as \say{high}, \say{medium}, or \say{low} difficulty area. Complementary, the operator can manually select the parameters for each area. Note that at the beginning of the simulation, the operator is required to set all these parameters (or confirm the provided default values).  
Finally, drones may have simulated malfunctions, in such cases, an alert is generated and requires the operator's attention.  



Overall, the set of actions that the operator can take, $A$, consists of five types of actions: 
(i) Changing the assumed probability of finding (additional) targets within any specific sub-area; (ii) Changing the area type of a given sub-area, denoted by $CT$; (iii) changing the scanning parameters associated with an area (i.e., altitude, velocity, and thresholds), denoted by $CP$;  (iv) Manually flying a drone that is suspected of being stuck, manually reporting on a target (without relevant alerts), and manually assigning a specific drone to a sub-area (v) Handling alerts such as detection, malfunctions, and intelligent messages (note that base on the information in the intelligent message the operator may decide to change the probabilities of the sub-areas or manually change the assignment of drones to sub-areas).

The simulation is high-resolution and realistic, incorporating advanced features such as simulated wind, dynamic movement of trees, and real-time imaging from drones, as can be seen in a short illustration video.\footnote{\url{https://youtu.be/W-HHF8s2O8c}}  

\subsection{The User Interface}
\begin{figure}[t]
\centering
\includegraphics[width=\columnwidth]{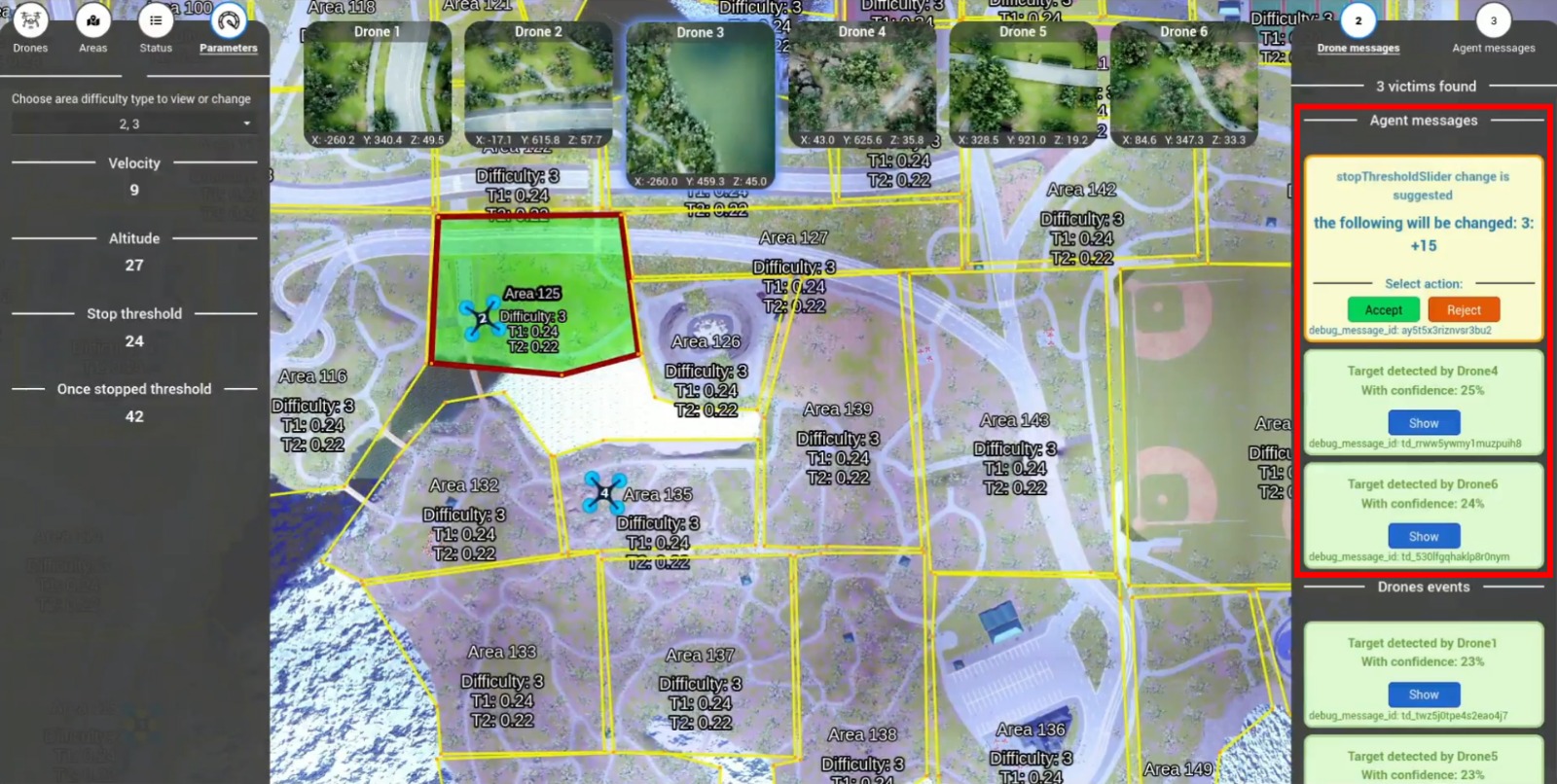} 
\caption{The SAR User Interface.}
\label{fig:sar_simulator}
\end{figure}

\begin{figure}[t]
    \centering
    \includegraphics[width=0.5\linewidth]{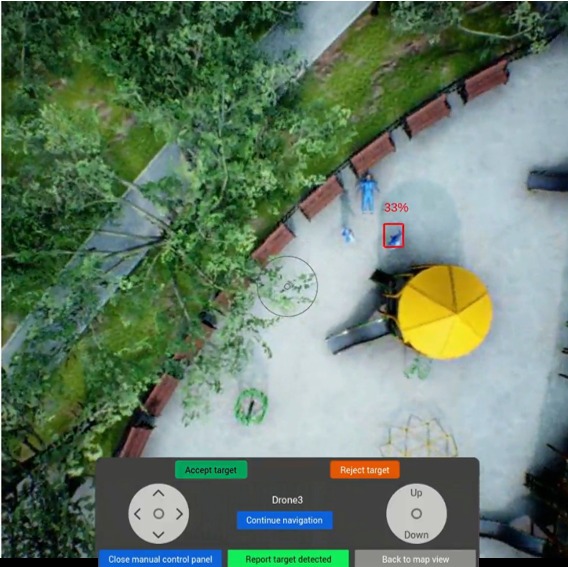}
    \caption{Manual control panel, during handling detection alert.}
    \label{fig:manual_control}
\end{figure}
\label{sec:sar}
\begin{figure}[t]
    \centering
    \includegraphics[width=\linewidth]{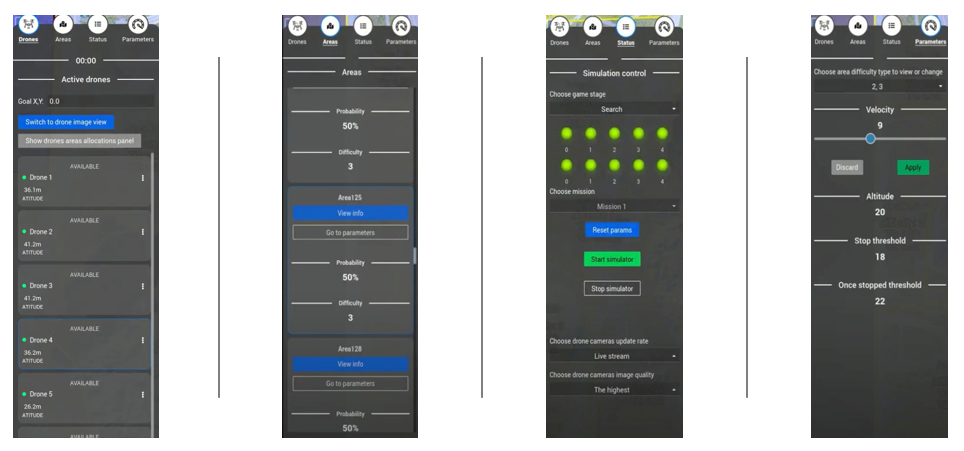}
    \caption{The left panel contain four tabs: Drones, Areas, Status and Parameters.}
    \label{fig:left_panel}
\end{figure}

Similar to the goal described by Chen et al. \cite{chen2022multi}, our SAR system aims to provide the operator with comprehensive situational awareness while minimizing the need for low-level control of each drone. The \say{task mode} includes automated allocation of the sub-areas to drones, where each drone scans its designated sub-area based on the difficulty level defined for that sub-area. The drones search for targets and alert the operator to suspected targets.
In the \say{command mode}, our system offers the flexibility for the operator to manually assign sub-areas to specific drones, manually control a drone, or manually report targets as necessary.

The central part of the user interface has two modes: a map mode and a drone mode. 
In the map mode, the map is displayed, divided into sub-areas, showing the locations of the drones on the map.  Each drone is labeled with a number, as recommended in~\cite{hoang2023challenges}.
At the top, there are small images for each drone displaying what the drone sees (see Figure~\ref{fig:sar_simulator}). In the drone mode, a specific drone's view is shown in at the central area in a larger format, allowing the operator to manually control the drone and consider the specific details observed by the drone. For example, the operator can better handle a detected target in this mode (see Figure~\ref{fig:manual_control}).

The left-side panel contains four tabs (see Figure~\ref{fig:left_panel}):
\begin{itemize}
    \item \textbf{Drones} - This tab provides each drone a list of its assigned sub-areas, with the option to manually change it.
    \item \textbf{Areas} - This tab displays all sub-areas, allowing the operator to change the probability of finding a target in a specific sub-area and amend the sub-areas difficulty levels.
    \item \textbf{Status} - Controlling the simulation mode - choosing a scenario and switching between scanning and parameters phases. 
    \item \textbf{Parameters} - In this tab, the operator can set the parameters (altitude, velocity, and thresholds) for each area.
\end{itemize}

On the right-hand side, there is a panel containing the alerts from drones, which are displayed in light green, and other messages, such as those from the agent, which are displayed in light orange (see Figure~\ref{fig:sar_simulator}). 

In a preliminary experiment, we separated these messages into two different tabs: drone and agent messages. However, we noticed that this setup was less convenient for most operators given the high number of alerts from the drones. In particular, operators noted that they felt pressured to respond to them and often missed important messages from the agent. Therefore, in our ensuing human evaluation, we combined both types of messages into a single tab, but with different colors, providing the operator with a simpler way to see and distinguish the alerts and agent's messages.

\section{Agent Implementation}

We instantiate the proposed advising agent  (Section \ref{sec:formal}) to the SAR task (Section \ref{sec:sar_mission}). To that end, we detail our implementation of the offline (reward estimation model) and online (advice provision) components.

\noindent{\bf\subsection{Reward Estimation Model}}
Next, we detail our implementation of Algorithm \ref{alg:generation}. 

For each $d\in D$, in order to select the initial state (Line \ref{S0} of Algorithm \ref{alg:generation}),  we use the parameters set by the operator at the beginning of the simulation and the initial assignment of drones to sub-areas.
For the $Pertuable(A_t,d,t)$ function, which varies the next action (Line \ref{Pertubate} of Algorithm \ref{alg:generation}) the action is selected through the following procedure: First, if in $d$ the operator performed an action of type $CP$ or $CT$ approximately at time $t$ (i.e., less than a minute away from $t$) then the action is selected. Otherwise, if $t=10min$, 
then we perform a random $CP$ or $CT$ change at probability $p = 1/|A'|$, where $A'$ is the set of actions we want to add at these points. Finally, we chose to handle one of the drone malfunction alerts (if it exists) by handling the highest confidence detection alert available (if it exists) or another alert at random. 

For $GetNewState(s_t,a_t)$ (Line \ref{GetNewState} of Algorithm \ref{alg:generation}), we simulate the effect of $a_t$ using our environment assuming it takes $\Tilde{C}_h(a) \forall{a\in A}$ time for the action to be executed. In our implementation, we define $\Tilde{C}_h(a)$ to be the average time it takes for the operator to perform an action $a$ in the entire trajectory. When handling a detection alert, we simulate whether the operator correctly or incorrectly handled it using her performance through the trajectory (i.e., using the true positive and true negative rates).
Note that since we run the simulator as in a real simulation, the changes in the state caused by the continued scanning of the drones are reflected in the next state in a natural manner. In other words, our synthetic simulation is restricted to automatically simulating the operator's actions.

For $QualityAssurance(D_{syn}, D, \Theta)$ and model training ($M$), we featurize the state-space such that each $s\in S$ is defined as follows:
\begin{itemize}[leftmargin=10pt]
\item Operator performance: 
$\Tilde{C}_h(a)$ that estimates the average time it took the operator to perform $a$ in the trajectory up to state $s$.  
\item Environment information: 
Remaining time for the SAR task, count of approved detection events, percentage of false detection alerts, count of current open alerts for each type of alert, lowest confidence thus far that led to detection, and count of actions performed by the operator.
\item Information for each area type:
The percentage of all sub-areas of that type that remained unsearched, the number of alerts generated from sub-areas of that type above and under current \detectiothreshold\ and  \pausetreshold\ until the current state, 
search parameters: \pausetreshold,  \detectiothreshold, \velocity, \altitude, and their last previous values (if exist). 
\end{itemize}

For $QualityAssurance(D_{syn}, D, \Theta)$, we first check the distance between synthetic trajectories and each of the real trajectories to ensure that the synthetic trajectories closely resemble the original trajectory they originated from.
Then, a clustering test of the synthetic and real trajectories is performed. This phase is used to ensure that each synthetic trajectory is assigned to the same cluster as the real trajectory and to make sure that the synthetic trajectories are roughly proportionally distributed among clusters. 
If the synthetic data quality assurance passes the required threshold, the algorithm trains a model. 
Then the quality assurance of the model is done through two perspectives: (1) Ensuring that the model accurately predicts the utility by leading to reasonably low MAE; and (2) Ensuring that the model brings about effective recommendations by aligning with actions known to perform well in hindsight through synthetic simulations (see \ref{subsub:UtilityFunction} for details).

\noindent\textbf{(1) Distances:}
First, we sample three states from each synthetic and real trajectory: one roughly at the beginning, one roughly at the middle, and one roughly at the end. We calculate the squared distance between the synthetic trajectory and each real trajectory in $D$ and sort them from lowest to highest. If the original trajectory from which the synthetic one originated is ranked high (i.e.,  first or second) in the sorted set, the synthetic trajectory can be considered adequate.  

\noindent\textbf{(2) Clustering:}
As before, we first sample three states from each synthetic and real trajectory: one roughly at the beginning, one roughly at the middle, and one roughly at the end. 
A K-means algorithm~\cite{jancey1966multidimensional, macqueen1967some,lloyd1982least, steinhaus1956division} was applied using both the real and synthetic trajectories. The proportion of synthetic trajectories that belong to the same cluster as their original one is returned (i.e., the higher the better).  

\noindent{\bf{Utility Function}}
\label{subsub:UtilityFunction}
Given a trajectory $\trajectory$ of length $l$ and $s_j \in \trajectory$, the discounted accumulated reward of a given state $s_j$ until the end of the trajectory is $\bar{R}(s_j)=\sum_{i=j,s_i \in \trajectory}^{i=l} \gamma^{i-j} R(s_i)$.
Since the detection of a target is a sparse event and most rewards along a trajectory are zero, we define a utility function $\UFunc$ that will be used to train a model $\Model$  that will enable the agent to compare states better. 

Let $\parametersForU=\{\paramForU_1,...\paramForU_m\}$ 
be a set of parameters of states in $S$.
We aim at developing a utility function $U(s)=\wights_0 \bar{R}(s)+\wights_1 \psi_1(s)+...+\wights_m \psi_m(s)$.
Given a state $s$ and actions $a$ and $a'$, let $\bar{s}$ and $\bar{s}'$ be the expected states generated by the expected state generator, respectively.
Our goal is to train a model $M$ to estimate $U$ such that: 

(I) if $M(\bar{s}) \geq M(\bar{s}')$ then $\bar{R}(s) \geq \bar{R}(s')$.

We considered the following parameters for $\parametersForU$:
(a) the path scanned by the drones until the end of the trajectory (i.e., to encourage scanning as much area as possible); (b) The average waiting time for an alert; (c)  The ratio between the number of detection alerts until the end of the trajectory and the number of alerts the operator can handle given his cost function (i.e., to encourage a balanced workload);  (d) The number of false negatives or false positives; (e) The number of targets correctly found until now in the trajectory; and (f)  The number of targets correctly found until the end of the trajectory.

To train a  model $M$ that attempts to satisfy condition  (I), the following three steps are performed  repeatedly:
\begin{enumerate}[nosep]
    \item Choose possible weights for $U$.
    \item Train a model $M$ that accurately estimates $U$.
    \item Evaluate the ability of $M$ to lead to effective action recommendations.
\end{enumerate}
Testing (2) is done by computing the MAE  (see \ref{modules_evaluation}). 
Testing (3), is done using a set of synthetic trajectories generated specifically for these tests (see \ref{sec:action_ordering}).
Once the parameters of $U$ have been established and $M$ has undergone training, the agent is presumably capable of generating effective actions for the operator.

\noindent{\bf\subsection{Advice Provision}}

Next, we detail our implementation of the action generator $(\AG)$ and the ranking model $(\RM)$.
The $\AG$ component uses the expected state generator to anticipate the outcome following the execution of an action, and then the reward estimation model, derived in the offline phase, to evaluate the effect of the action in terms of expected reward.
Every predetermined number of seconds, the agent activates the $\AG$ and the $\RM$ in order to produce $\adviceNum$ actions that are expected to be the most promising given the current state. In addition, every incoming alert from the drones activates the $\RM$ once again in order to determine whether the new alert should result in a different ranking. 

The process of generating possible actions involves two types of actions (corresponding to the actions available in the system): First, an action that is not associated with an alert generated by the system, i.e.,
(i) Changing the area type of a given sub-area, denoted by $CT$; (ii) changing the scanning parameters associated with an area (i.e., altitude, velocity, and thresholds), denoted by $CP$; (iii) Handling intelligent messages;  (iv) Detecting a drone that is not progressing and may be stuck (and has not sent an alarm).
The second type of actions are actions that are directly associated with alerts generated by the system, i.e., Handling alerts such as detection, with low-confidence (\pausetreshold) or with high-confidence (\detectiothreshold) and the malfunction of a drone. The action generator uses the expected state generator to generate the expected state and then uses the reward estimation model. 

Advice can be a simple action, e.g., increasing the parameter $\pausetreshold$ by $5$ for area type \say{low} or changing the area type of a given sub-area to \say{high}. In our implementation, we further allow for complex actions to be proposed, e.g., a list of changes for the same parameter for each area type or a list of changes of the area type for several sub-areas. 
Generating and evaluating complex actions leads to a combinatorial challenge.
Hence, a two-phase algorithm is proposed: initially, simple actions are generated and assessed; subsequently, the most promising ones are combined into complex actions, $a_i^v\subset A$.
For example, consider changing the parameter $\pausetreshold$.  
In the first phase, for each area type, we consider changing $\pausetreshold$ by    $\pm0.5x, \pm x$ and $\pm2x$ where $x$ denotes a predefined value or the mean change that the operators performed for $\pausetreshold$ in $D$. The expected state generator is then used for these possible changes followed by the reward estimation model. 
In the second phase, the agent considers adjusting the $\pausetreshold$ values for all area types by modifying the existing $\pausetreshold$ values according to the result of the first phase. the best value identified in the initial phase. The action generator uses the expected state generator in order to generate the expected state if these changes were to be performed considering the expected time it would take for the operator to change these parameters. 



We rank the actions provided by the action generator according to their estimated reward and provide the three top-ranking ones to the operator (negatively estimated actions are not provided). Tie-breaking favors the handling of detection alerts over others and alerts with higher confidence over others as a secondary criterion. 
The operator has the option to perform an action that is not advised by the agent. It is important to note that some actions are typically based on information the agent cannot access, such as changing the probability of a sub-area, which requires interpreting intelligence messages and understanding the map. Therefore, the operator may decide to perform such actions independently.



In our implementation, the expected state generator component is designed to predict the next state in a way that reflects the main significance of the action. The \textbf{operator performance} remains unchanged. The \textbf{environmental information} is modified as follows: the remaining time for the task is updated based on $\Tilde{C}_h(a)$; the count of approved detection events is adjusted if the action handles a detection alert, reflecting the percentage of approved events up to the current state; the percentage of false alerts remains unchanged; the current open alerts decrease by one if the action handles a detection alert; the lowest confidence that led to detection remains unchanged; and the count of actions performed by the operator increases by one, depending on the action type.
\textbf{Information for each area type} is also considered: the percentage of unsearched sub-areas remains unchanged, the number of alerts generated from sub-areas above and below the current \detectiothreshold\ and \pausetreshold\ is updated if the action changes these thresholds, and the search parameters are modified if the action alters one of them, with the previous values also being updated accordingly.
Note that changes in state due to the continued scanning of drones are not included as the exact computation is highly time-consuming. 

\section{Evaluations}

Recall that the development of our advising agent relies primarily on two components. The offline:  training the reward estimation model.
The online: developing and testing the advising agent uses the reward estimation model in real-time to provide advice during the simulation. 
Next, we discuss the details of the simulation used in the user study, followed by two experiments. The first experiment was conducted to collect data necessary for training the model, while the second experiment aimed to evaluate the quality of the advice provided by the agent during the simulation.

\noindent\textbf{Simulation.}
We designed and used a simulated drone-based SAR environment within Microsoft's AirSim based on the SAR task definition provided in Section \ref{sec:sar_mission}. The environment features six drones and a large search zone resembling a town with buildings, roads, parks, and playgrounds. This town is based on the known unreal-engine map called \say{City Park} with small modifications (https://www.unrealengine.com/marketplace/en-US/product/city-park-environment-collection). 
Half of the search area was used for Experiment 1, whereas the other half was used for Experiment 2. In both cases, the entire map was divided into $25-50$ smaller sub-areas. 

\noindent\textbf{Participants.}
Overall, forty subjects participated in our experiments and were recruited in two batches: the first batch performed the SAR task without the assistance of the agent, and the second batch was asked to perform the SAR task with and without the agent (on different scenarios).
The participants ages range from $18$ to $62$ (mean $28$), $65\%$ males, and $35\%$ females. Most participants work in the tech industry or are Computer Science students with some experience in video games (a prerequisite for participation). Each subject in the first batch was paid \$26 and this batch was used for training.  In the second batch, each participant received a base payment of \$16 for their time and an additional 15 cents for each target they identified. This incentive was designed to encourage maximum effort during the simulations, where the primary evaluation metric was the number of targets identified.


\noindent\textit{Training:}
Before engaging in the SAR task, each operator received a hands-on tutorial to familiarize themselves with the system. Then, at the beginning of the simulation, the operator received an initial intelligent message detailing the background story (e.g., a lost group of travelers). Then, the operator had $5$ minutes to configure the system as detailed in Section \ref{sec:sar_mission}, followed by 20 minutes to perform the task (i.e., to search and find as many desired targets as possible). 
Overall, four different scenarios, each encompassing a different number of targets (18-20), their locations, and the content of two intelligence messages were manually designed by the authors and explored through preliminary testing to ensure varying conditions. 

\subsection{Experiment 1: Reward Estimation}
In this experiment, we used machine-learning techniques to estimate the long-term reward of performing an action at a given state. We examined whether the ordering over actions, induced by the reward estimation model, is correct by simulating the long-term effects of different actions and we evaluated the quality of the synthetic trajectories.
To focus on high-quality demonstrations, we considered the trajectories of operators of the first batch of subjects who had found at least one target during the simulation and denoted this set as $D$. 

\paragraph{\bf Machine Learning Performance}
\label{modules_evaluation}
The demonstrated trajectories, $D$, were used as input to Algorithm \ref{alg:generation}, resulting in approximately 1000 synthetic trajectories. 
We used two supervised learning models: a Random Forest (RF) \cite{randomForest} and a Long Short-Term Memory (LSTM) \cite{LSTM} to estimate the long-term reward of performing an action at a given state. To evaluate the models' accuracy, we used an 80-20\% train-test split and calculated the Mean Absolute Error (MAE). 
Overall, Random Forest provided a notably lower MAE than the LSTM ($0.55$ vs $1.05$).
Given that the utility function ranges from 1 to 22, the $0.55$ MAE demonstrates the model's effectiveness in supporting the selection of the appropriate recommendation action.


\paragraph{\bf Ranking Model}
\label{sec:action_ordering}
We examined whether the estimation model leads to good action recommendations. We focused on the $CP$ action for $\detectiothreshold$ and $\pausetreshold$ since it had been the most challenging action to evaluate and rank and had rarely been performed by human operators without the assisting agent. To simulate the long-term effect of performing the $CP$ action for $\detectiothreshold$ and $\pausetreshold$,  for each $d\in D$, we generated $30$ new synthetic trajectories using the same procedure as in Algorithm \ref{alg:generation}.
In the third of them, the Pertuable function did not perform $CP$. In another third of them, a $CP$ action that increased $\detectiothreshold$ and $\pausetreshold$ was performed in the middle of the trajectory, and in the third, a $CP$ action that decreased $\detectiothreshold$ and $\pausetreshold$ was performed also in the middle of the trajectory.

For each $d \in D$, we computed the average number of targets found in each type of synthetic trajectory, resulting in three averages per $d$: one for synthetic trajectories without the $CP$ action, one for synthetic trajectories with the $CP$ action that increases thresholds, and one for synthetic trajectories with the $CP$ action that decreases thresholds. We then checked if our model gave the highest rank to the action that led to the highest average among these three possibilities.
Then, for each $d \in D$, we trained a model with the same utility function, but in which the synthetic trajectories of $d$ were in the test set, and used this model to check if the action that received the highest rank was indeed the one corresponding to the synthetic trajectories where the highest average number of targets was found.
\begin{figure}[tbp]
    \centering
    \includegraphics[width=0.75\columnwidth]{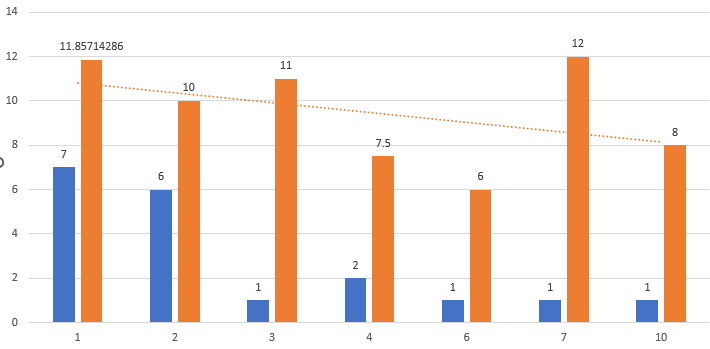}
    
    \caption{Orange bars are the average number of true targets that were approved. Blue bars are the number of simulations that were performed in this category.}
    \label{fig:stuck-half-label}

\end{figure}
\noindent{\it Results.}
\begin{itemize}[leftmargin=10pt,topsep=0pt]
    \item For three $d\in D$, the highest average number of targets found was in the synthetic trajectories generated without performing $CP$. In these cases, our model correctly gave the highest rank for handling detection alert action and avoided recommending $CP$.
\item For five  $d\in D$, the highest average number of targets found was in the synthetic trajectories generated by performing the $CP$ action that increases $\detectiothreshold$ and $\pausetreshold$.
Within $11\%$ of the states, the recommendation derived from model $M$ accurately advised an increase of the detection threshold, while in $89\%$ of the states, the recommendation suggested handling a detection alert. 
Notably, the model correctly did not advise decreasing the threshold.
    
    \item  There were no  $d\in D$, such that the highest average performance in the synthetic trajectories was of the  synthetic trajectories generated by performing the $CP$ action that decreased $\detectiothreshold$ and $\pausetreshold$
\end{itemize}
The results highlight the robustness of our model, as it consistently avoids generating recommendations that could be proven incorrect. Additionally, since human operators rarely performed $CP$ actions in scenarios without the agent's guidance, the agent’s cautious approach to recommending these actions proved advantageous, as demonstrated in Experiment 2 (section \ref{EX2}). This cautious strategy resulted in a high acceptance rate of the agent’s recommendations by the participants, leading to a significant improvement in team performance.

\paragraph{Handling Malfunction Alert Prioritization.}
Recall that our ranking model favored handling drone malfunction alerts. We validated this heuristic based on the demonstrations from human operators $D$ examining the relation between the time that passed between the appearance of a drone malfunction alert and the time that the operator handled the alert.

\textit{Results:}

Figure \ref{fig:stuck-half-label} displays the average number of targets found, divided into bins based on the time that passed between the appearance of a drone malfunction alert and the time that the operator handled this alert (orange), and the number of human operators present in these bins (blue). The results show that the simulations in which the operator waited before handling the stuck drone alert typically resulted in fewer targets found supporting our heuristic choice.


\paragraph{\bf Quality Assurance}
For this evaluation, we used the second batch. We chose one of the scenarios and collected the trajectories of the human subjects that did not get the agent's assistance. This yielded ten trajectories.
We generated $6$ synthetic trajectories for quality assurance for each real one. These trajectories were analyzed in two ways: using a distance-based approach and using a clustering analysis approach.  

\noindent\textbf{(1) Distance:} 
Starting with the distance-based analysis, we examined the distances between the synthetic and real trajectories of each operator in two ways: First, for each operator, we examined whether the synthetic trajectories indeed best resembled the original using a standard distance measure. Second, for each operator, we calculated the average distance of the six synthetic trajectories to each one of the real trajectories and normalized it by dividing it by the sum of the average distances of the synthetic trajectories to all real trajectories.
Thus, for each operator, the sum of the normalized distances from the synthetic trajectories to the real trajectories was $1$. Using statistical testing, we examined whether the synthetic trajectories better resemble the originating one than any other real trajectory.  

\begin{figure}[tbp]
    \centering
    \includegraphics[width=\columnwidth]{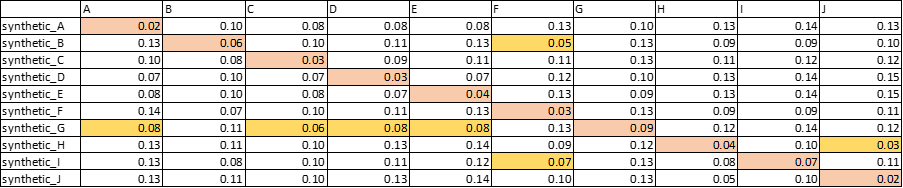}
    \vspace{-10pt}
    \caption{Normalized distances between real trajectories $(A-J)$ and the averaged synthetic trajectories ($synthetic\_A-synthetic\_J$.)}
    \label{fig:normlized_distances_paper}
\end{figure}

\textit{Results:}
For $6$ out of the ten operators, all $6$ synthetic trajectories were the most similar to the real trajectory of their operator (see Figure~\ref{fig:normlized_distances_paper}.
For $3$ additional operators, all $6$ synthetic trajectories were second most similar to their operator's real trajectory.
 
The maximum normalized distance from the averaged synthetic trajectories to the real trajectory of the same operator was $0.09$ and the average was $0.04$. That is, for each operator, the distance between her synthetic trajectories and her real trajectory was lower than the average distance between her synthetic trajectories and any real trajectory (i.e., $0.1$). 

\noindent\textbf{(2) Clustering:} We used a clustering-based analysis with a K-means algorithm applied to all trajectories alike, subject to the condition that each real trajectory should be assigned to a different cluster. Then, we examined the portion of synthetic trajectories that were assigned to the same cluster as their originating one.   
\textit{Results:}
For 58 of the 60 synthetic trajectories, the synthetic trajectories were assigned to the same cluster as their originating real ones.  The remaining two synthetic trajectories originated from a single human operator.
Based on the evaluations using the two measurements, we concluded that our method for generating synthetic trajectories from human trajectories successfully produced trajectories that closely resembled the original ones.

\subsection{Experiment 2: Advising Agent}
\label{EX2}
We tested three hypotheses stated below.
\begin{itemize}[leftmargin=10pt,topsep=0pt]
 \item {\bf H1} The advising agent improves the team's performance as measured by the number of targets found and the covered area.
\item {\bf H2} The human operators are satisfied by the advising agent, improving their perceived ability to manage more drones effectively. 
\end{itemize}
While the agent guides the human operator on which actions to take, it also introduces additional tasks alongside those generated by the drones. The following hypothesis proposes that these additional tasks do not increase the operator's cognitive load, as the benefits of the agent's advice on the action to take next, effectively offset the added demands.
\begin{itemize}[leftmargin=10pt,topsep=0pt]
\item {\bf H3} The agent's assistance does not substantially alter the mental demands on the human operators.
\end{itemize}

\noindent\textbf{Experimental Procedure.}
Human operators from the second batch participating in the study completed four scenarios in a fixed order. Half of them first completed the initial two scenarios with the help of our agent, followed by two scenarios without it. The other half started with two scenarios without our agent, followed by two with its help. 
The experiment was conducted in two sessions, each of two scenarios, to mitigate fatigue. After each session, operators filled out a standard NASA-TLX questionnaire \cite{Hart1988} to assess workload. Upon completing both sessions, a final questionnaire regarding the entire experiment was administered.

\noindent\textbf{Data processing.}
The recorded sessions were analyzed using three key analyses: First, we statistically evaluated the accumulated rewards and other performance metrics (e.g., the area covered) of the human-multidrone team performance with and without our agent's assistance. Second, we considered the questionnaire reporting.  
Last, we statistically evaluated the cognitive load on the operator with and without our agent's assistance.

\begin{figure}[htbp]
\centering
\includegraphics[width=0.9\columnwidth]{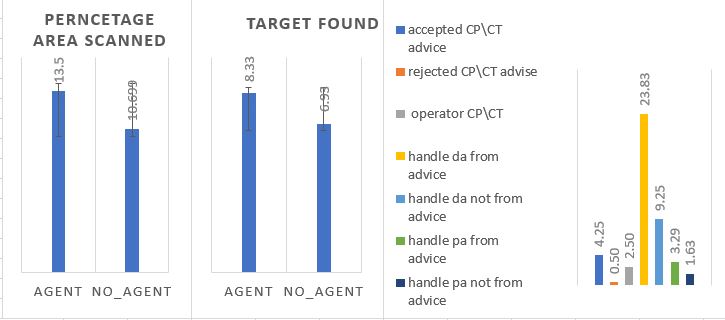} 
\vspace{-10pt}
\caption{ \small The first two graphs present the performance with the help of the agent and without it. The third graph presents agent-human interaction within missions. }
\label{fig:results_graphs}
\end{figure}

\noindent\textbf{Results.}
As illustrated in Figure \ref{fig:results_graphs}, we observed a statistically significant improvement in the number of targets found, with a $20\%$ increase ($p=0.04$, via one-sided t-test), as well as a significant $26.2\%$ enhancement in the percentage of the search area scanned ($p=0.002$, via one-sided t-test). {\it Thus, the data supports H1.}
This could be explained by observing the acceptance of the agent's generated advice, our analysis indicates that, in most instances, the operators followed the agent's recommendations.
We observed a statistically significant increase in the number of $CP$ or $CT$ actions ($p=0.0004$, via one-sided t-test). In scenarios where the advising agent was present, operators executed actions like $CP$ or $CT$ an average of $6.75$ times, with $4.25$ (blue column) of those actions being a direct result of the agent's advice. In contrast, operators performed such actions only $2.25$ times on average when our agent was not present.  

When observing the handling of the detection alerts, when the agent was present, operators preferred handling the alerts coming from the agent (yellow column) over alerts that were not in the advice section (turquoise column). We observed a significant number of detection alerts handled outside of the advice section. This can be explained by the time lag required for updating the advice list (a few seconds due to calculations). 

\begin{figure}[htbp]
\centering
\includegraphics[width=0.9\columnwidth]{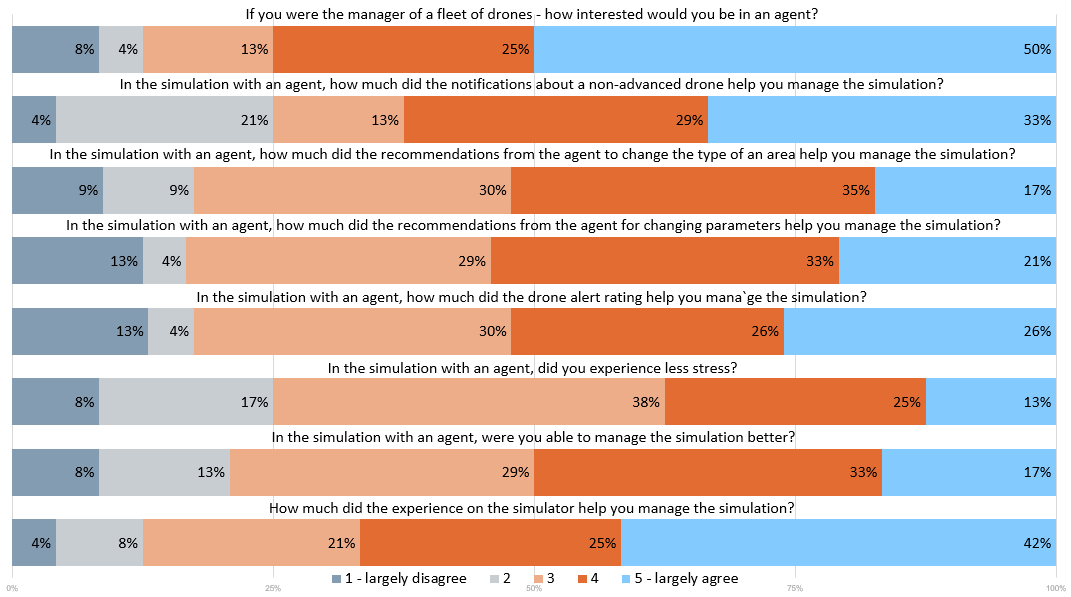} 
\vspace{-10pt}
\caption{ \small A survey on SAR task with the advising agent}
\label{fig:sar_survey}
\end{figure}

The questionnaire results (Figure \ref{fig:sar_survey})
suggested a strong preference among the participants for the advising agent.
In addition, most participants expressed a preference for using the agent in SAR tasks, confidently stating that it would enable them to handle more drones successfully. Specifically, the average number of drones human operators believed they could manage with the agent was $7.54$, compared to $6.83$ without the agent ($p=0.03865$ via one-sided t-test). We observed that the operators were generally pleased with the agent's advice regarding $CP$, $CT$, and stuck drone messages. Overall, the participants in our study presented favorable perceptions of the agent. {\it Thus, the data supports H2.}

Considering the cognitive load, the TLX scores indicate that the overall cognitive load was not statistically different between the two setups, although it was slightly lower when the agent was present ($57.52$ with the agent vs. $58.94$ without the agent, see Table~\ref{tab:tlx} for details). Also, the questioner's results were aligned with the NASA-TLX results; it was evident that the agent did not introduce any additional cognitive load to the participants' perceived stress levels.
{\it Thus, the data supports H3.}

\begin{table}[]
    \centering
    \begin{tabular}{c|c|c}
         & Agent & No agent \\
        Effort & 60.11, 3.44 & 61.35, 3.38 \\
        Frustration & 46.82, 1.94 & 45.97, 1.61 \\
        Mental demand & 58.23, 2.94 & 60.14, 3.02 \\
        Performance & 48.67, 3.67 & 54.55, 4.41 \\
        Physical demand & 28.44, 0.52 & 22.26, 0.79 \\
        Temporal demand & 55.14, 2.47 &  54.94, 2.76 \\
    \end{tabular}
    \caption{NASA-TLX, the first number is the rating, the second is its weight in the overall score calculation.}
    \label{tab:tlx}
\end{table}
\section{Conclusions}

Operators play a pivotal role in multi-drone systems, particularly in collaborative missions such as Search and Rescue (SAR) which require human supervision and control. In this article, we presented an advising agent methodology that we implemented and evaluated extensively with human participants in a simulated multi-drone SAR setting. Our results suggest that human-multi-drone team performance can be significantly improved through the introduction of the advising agent. In order to effectively generate the advice, the results further show that closely mimicking human operators' actions is possible and requires only a limited number of demonstrations. Additionally, the introduction of the advising agent does not seem to bring about any increase in the operators' cognitive load (if any, it seems to slightly reduce it),  and participants expressed a strong preference for the use of the advising agent while asserting that it would enable them to effectively manage a larger number of drones. Taken jointly, our results seem to suggest that the integration of advising agents in complex human-in-the-loop systems, and especially human-multi-drone teams, may be advantageous and may be achieved with only limited human demonstrations.

\bibliographystyle{acm}
\bibliography{mybibfile}

\appendix
\section*{A1. Data Collection with Human Operator}
For the purpose of training data collection, we delineated a set of feasible tasks for human operators and extracted relevant telemetry from these simulations, to subsequently be utilized by the AA agent. The following list enumerates all potential operator tasks:

\begin{itemize}
    \item Define the area type for each sub-area.
    \item Specify the probabilities of human presence in each area.
    \item Adjust existing area probabilities or difficulty levels.
    \item Prescribe altitude, velocity, thresholds: low-confidence (lc) and high-confidence (hc) corresponding to each area's type.
    \item Approve or reject detection alert.
    \item Manually operate drones.
    \item Handle malfunction alert.
    \item Manually detect humans - press a stop buttons and enforcing target acceptance.
    \item Manually modify the assignment of drones to sub-areas.
    \item Direct focus to specific cameras.
\end{itemize}

From each simulation with human operator we collect the following data:
\begin{itemize}
    \item The sequence of actions performed by the operator, including their timing.
    \item The temporal allocation of drones to specific sub-areas.
    \item The entire world's state each time.
    \item The drones' position at each time.
    \item Dashboards' screens, where user interaction was most pronounced.
    \item NASA Task Load Index (TLX) assessments conducted post-experiment.
\end{itemize}

\section*{A2. Mimicking Human Decision Making}
\label{SimulationRuns_appendix}
As discussed in Section $4.2$ of the paper, acquiring a substantial amount of data is crucial for developing a action generator and a ranking model that can mimic the decision-making trajectory of a human operator. In our simulation platform, we conducted simulations involving human operators, excluding the Advising Agent, to collect all relevant information for $QualityAssurance(D_{syn}, D, \Theta)$. 
After analyzing the collected data, we identified several essential parameters necessary for accurately mimicking the original operator's behavior:
\begin{itemize}
    \item Same initial probability for each search sub-area.
    \item Same area type that the operator assigned to each search sub-area.
    \item Scanning order of the search sub-area and allocation for each drone that followed the operator's real simulation (if the operator did not complete scanning all search sub-areas, we allocated them according to the probability of the areas in descending order).
    \item Adjustable parameters for each difficulty type (detection threshold, alert threshold, altitude, velocity).
    \item Same changing parameter actions that the operator took at the same timestamp as the actual simulation.
    \item True Positive (TP) rate - the probability of approving a detection when a target is in the drone's camera view.
    \item True Negative (TN) rate - the probability of rejecting a detection when a target is not in the drone's camera view.
    \item The average time for handling a detection alert.
    \item The average time for handling $CP$ action.
\end{itemize}

\section*{A3. Operator Post-Scenarios Questionnaire}
 after finishing the four scenarios, each human operator filled out a detailed questionnaire.  Their answers were on a scale from $1$ to $5$.
 The questionnaire had four main parts, asking about different aspects of the operators' experiences.
\begin{itemize}
    \item  Personal Information - 
        \begin{itemize}
            \item Name
            \item Gender
            \item Age
            \item Occupation
        \end{itemize}

    \item  Background - 
        \begin{itemize}
            \item Have you flown a real drone before?
            \item Have you photographed with a drone before?
            \item How often do you play computer games?
            \item Have you played a computer game that simulates flying an aircraft?
        \end{itemize}
    \item  Post-Game Questions - 
        \begin{itemize}
            \item How comfortable was the interface for giving probability to areas?
            \item How comfortable was the interface for selecting the parameters for each area type?
            \item To what extent did the alerts from the drones help manage the simulation?
            \item To what extent did you feel that her actions were consistent with finding missing persons?
            \item How much did the experience on the simulator help you manage the simulation?
            \item To what extent did the explanations that appeared on the alert help you manage the simulation?
            \item How much did the manual driving option help you manage the simulation?
            \item How many drones do you think can be managed without an agent in a good way?
        \end{itemize}
    \item Post-Game Agent Questions - 
        \begin{itemize}
            \item In the simulation with an agent, were you able to manage the simulation better?
            \item In the simulation with an agent, did you experience less stress?
            \item How many drones do you think can be managed with an agent in a good way?
            \item In the simulation with an agent, how much did the drone alert rating help you manage the simulation?
            \item In the simulation with an agent, how much did the recommendations from the agent for changing parameters help you manage the simulation?
            \item In the simulation with an agent, how much did the recommendations from the agent to change the type of an area help you manage the simulation?
            \item In the simulation with an agent, how much did the notifications about a non-advanced drone help you manage the simulation?
            \item If you were the manager of a fleet of drones - how interested would you be in an agent?
        \end{itemize}
\end{itemize}


\section*{A4. Centralized Real-Time Task Allocation Algorithm}
\label{CentralizedAlgorithm}

We used a genetic algorithm for path planning suggested in \cite{du2019evolutionary}. The algorithm was originally used for finding lost tourists in large national parks in China. According to the suggested algorithm, a team of drones is looking for a target based on probabilities of the target location and image analysis. This strategy made the algorithm a good choice for our use.

The algorithm's input is:
\begin{itemize}
    \item A - area
    \item Set of n drones $\{drone_1,...drone_n\}$
    \item $m$ sub-regions ${a_1, a_2,..., a_m}$ based on the topographic features.
    \item $a_j0$ - the initial location of $drone_j$
    \item $d_i$ - the distance from $a_{i0}$ to sub-area $a_i$
    \item $T$ - the maximum allowable time of the operation. 
    \item K - the number of search modes of the drones (the possible search altitudes)
    \item a prior probability  $p_t(i)$ of target location in each sub-area $a_i$ at time t $(i\in\{1...m\}, t\in\{1...T\})$  
    \item $\Delta T (i, j, k) $ - time required by drone $drone_j$ to search sub-area $a_i$ with mode k
    \item $\Delta t(i,i',j)$ - the time for drone $drone_j$  to fly from a sub-area $a_i$ to another $a_i'$.
\end{itemize}

The algorithm’s output is a search path $x_j$ of each drone $drone_j$ such that the object can be detected as early as possible.

 $x_j=\{(a_{j,1}, k_{j,1}), (a_{j,2},k_{j,2}), ... , (a_j,m_j, k_j, m_j)\}$, where $\{a_{j,1}, a_{j,2}, ... , a_j,m_j \}$ is the sequence of sub-areas to be searched by $drone_j$ , and $k_{j,i}$ is the search mode used for the $i_{th}$ sub-area $a_{j,i} (1 \leq i \leq m_j)$.

Based on the search path $x_j$, the search times $\Delta \tau (i, j, k)$, and the flight times $\Delta t(i, i', j)$, the action of drone $drone_j$ at each time $t$ can be determined.
$x_t(j)=(i,k)$ are used to denote that $drone_j$ is searching in sub-area $a_i$ with mode k, and $x_t(j)=(i, i')^T$  to denote that $drone_j$ is flying from a sub-area $a_i$ to another $a_i'$ .

The objective function for the optimizations can be calculated in the following way:
Let $t^*$ be a hypothetical time at which the target is detected. iteratively the probability of $t^*=t$ is computed for all $t$ since the occurrences of detection by separate drones may be considered mutually exclusive.

$P(t^*=0)=0$

$P(t^*=t)=P(t^*=t \mid t^*\geq t)P(t^*=t)=[\sum_{i=1}^m\sum_{j=1}^n\sum_{k=1}^K{p_t (i) p_t (i,j,k \mid x_t (j)}] \times [1-\sum_{t'=0}^{t-1} P(t^*=t')]$

$p_t(i,j,k \mid x_t(j))=$ 
\[\begin{cases}
    (p_t (i,j,k),& \text{if }  x_t (j)=(i,k)\\
    0,              & \text{otherwise}
\end{cases} \]

The time complexity of the objective function $O(mnKT^2 )$.

A primary technique for growing a population of possible solutions to the problem and a sub-procedure for optimizing each drone path in the solution make up the genetic algorithm.

First, it initializes a population of $N$ solutions, including $N-1$ randomly generated solutions and a solution of a greedy method that selects the next step with the highest payoff (in terms of the ratio of the detection probability to the time consumed).
Then do the following steps until time has terminated:
For each solution $X$ and for each path $x_j$ in X, the sub-algorithm is used to optimize $x_j$ and evaluate the objective function for the solution X. After evaluating all objective functions, for each solution $\lambda(X)$, which denotes the migration rate of solution X, will be calculated and then migration or mutation will be performed on the solution.

The sub-procedure mentioned above for path optimization is as follows. Suppose a set $C_j \subset A$ of sub-areas have been assigned to drone $drone_j$. The sub-procedure produces the search path $x_j$ of $drone_j$ based on the NEH heuristic \cite{nawaz1983heuristic} and tabu search method \cite{glover1989tabu, glover1990tabu}.
The ratio of the overall detection probability to the entire time spent along the path is used to assess the fitness of a path $x_j$. The function is described as the probability that the $j_{th}$ drone search sub-area $a_i$  times probability at time $t$, the target will be at sub-area $i$ searched by drone $j$ with mode $k$, divided by the time required by drone $drone_j$ to search sub-area $a_i$ with mode $k$ for all $i$ in path plus flying time from each sub-area. 
The migration algorithm is based on the BBO metaheuristic (Biogeography-Based Optimization), and it allows a solution to migrate characteristics from other solutions.
For each solution in the population, a nonlinear model is used to calculate its migration rate $\lambda (X)$ as - $\lambda (X)=0.5-0.5cos(\frac{f(X)-f_{min}+\epsilon}{f_{max}-f_{min}+\epsilon} \pi)$
$f_max$ and $f_min$ are the population's highest and minimum objective function values.
In each generation, each solution X has a probability $\lambda(X)$ of importing characteristics from other solutions, which are chosen from the top half of the population with probabilities proportionate to their fitness.
A solution that is not migrated will be mutated by regenerating the search paths for some drones. Each solution X is assigned a mutation rate $\mu(X)$, initialized to 0.5 and updated at each generation as: $\mu(X)=\mu(X)\cdot \alpha^{-\frac{f(X)-f_{min}+\epsilon}{f_{max}-f_{min}+\epsilon}}$. The mutation randomly assigns a sub-area to drones which are selected for mutation. If a solution has not improved after {\it g} (a control parameter commonly set to 6) generations, it will be replaced with a new solution generated randomly to increase solution variety.

\end{document}